Entire latex paper,  no figures,  revised,  to appear in Phys. Lett. B
\documentstyle[prl,aps]{revtex}
\begin{document}
\draft
\title{Hadronic matter compressibility \\ from \\
event-by-event analysis of heavy-ion collisions}

\author{Stanis\l aw Mr\' owczy\' nski\footnote{Electronic address:
mrow@fuw.edu.pl}}

\address{So\l tan Institute for Nuclear Studies,\\
ul. Ho\.za 69, PL - 00-681 Warsaw, Poland\\
and Institute of Physics, Pedagogical University,\\
ul. Konopnickiej 15, PL - 25-406 Kielce, Poland}

\date{8-th December 1997, Revised 22-nd February 1998}

\maketitle

\begin{abstract}

We propose a method to measure the hadronic matter compressibility 
by means of the event-by-event analysis of heavy-ion collisions 
at high energies. The method, which utilizes the thermodynamical
relation between the compressibility and the particle number fluctuations,
requires a simultaneous measurement of the particle source size, 
temperature and particle multiplicity. 

\end{abstract}


\vspace{0.5cm}
 

\vspace{0.5cm}


Large acceptance detectors allows one for a detailed analysis of 
individual collisions of heavy-ions at high-energies. Due to hundreds
or even thousands of particles produced in these collisions, variety of
powerful statistical methods can be applied. Then, such event-by-event 
studies can provide valuable dynamical information which is otherwise
hardly available. For example, we have shown in \cite{Gaz92}, see also
\cite{Gaz97}, that the correlation between particle multiplicity and their 
average transverse momentum, which is observed in proton-proton interactions, 
causes sizeable fluctuations of the total transverse momentum of particles 
produced in a single nucleus-nucleus collision if such a collision is a 
superposition of independent nucleon-nucleon interactions. The 
fluctuations from Pb-Pb collisions at 158 GeV per nucleon studied in 
NA49 experiment have appeared to be noticeably smaller than the properly
normalized fluctuations from the proton-proton interactions at the same
collision energy \cite{Rol97}. Thus, a substantial role of the secondary 
interactions in heavy-ion collisions has been proven without a reference 
to any collision model.

Stodolsky \cite{Sto95} and Shuryak \cite{Shu97} have made another 
interesting proposal of the event-by-event analysis. They have adopted 
a standard assumption that the hadron matter from the nuclear collisions 
is in thermodynamical equilibrium. Then, according to the well known 
thermodynamical relation, the temperature fluctuations can be treated as 
a measure of the heat capacity of the hadronic matter. Shuryak \cite{Shu97} 
has also briefly considered the relation, which couples the particle number 
fluctuations to the particle number derivative with respect to the chemical 
potential. In this paper we follow a similar line of reasoning and present 
a method to determine the hadronic matter compressibility via the multiplicity 
fluctuations. 

The method is based on the thermodynamical relation, see e.g. 
\cite{Hua63,Lan80}, which expresses the dispersion of the particle 
number $N$ in a volume $V$ through the isothermal compressibility as 
\begin{equation}\label{compres}
{\langle N \rangle^2 \over \langle N^2 \rangle - \langle N \rangle^2}
= - \; {V^2 \over T}\;
\bigg({\partial p \over \partial V}\bigg)_{T,\langle N \rangle} \;,
\end{equation}
where $T$ is the temperature and $p$ the pressure. In principle, the
relation (\ref{compres}) allows one to find $\partial p/\partial V$ as
a function of $\langle N \rangle$, $V$ and $T$ and then to reconstruct
the equation of state. Below we discuss how to realize this program 
in heavy-ion reactions.

There are many sorts of hadrons 
$(\pi^+, \: \pi^0, \: \pi^-, \: K^0, \: \bar K^0, .....\:)$
in the final state of nuclear collisions. Therefore, our first task 
is to generalize eq.~(\ref{compres}) to the system of $k$ components. 
To solve this problem one has to assume that the hadron system at
freeze-out is not only in the thermal but in the chemical equilibrium 
as well. The distinction between the thermal and the chemical freeze-out
is discussed at the end of this paper. The extensive analysis 
\cite{Bra95,Bec97} of the experimental data from AGS and SPS accelerators 
shows that the assumption of the chemical equilibrium is very well satisfied 
in heavy-ion collisions at least for nonstrange particles.

To derive the relation analogous to (\ref{compres}) we follow \cite{Hua63} 
and write the grand canonical sum for the $k-$component system as 
$$
\Xi(z_1,...\:z_k,V,T) = \sum_{N_1,...\:N_k} 
z_1^{N_1}...\:z_k^{N_k} \; Q_{N_1,...\:N_k}(V,T) \;,
$$
where $z_i$ is the fugacity of the $i-$th component and 
$Q_{N_1,...\:N_k}(V,T)$ is the canonical partition function.
Keeping in mind that 
$$
Q_{N_1,...\:N_k}(V,T) = e^{-\beta F(N_1,...\:N_k,V,T)}
$$
with $\beta \equiv T^{-1}$ and $F(N_1,...\:N_k,V,T)$ being the system
free energy, one finds that the probability $P_{N_1,...\:N_k}$ to find
$\{ N_1,...\:N_k \}$ particles is proportional to 
$$
P_{N_1,...\:N_k} \sim z_1^{N_1}...\:z_k^{N_k} \; 
e^{-\beta F(N_1,...\:N_k,V,T)} \;.
$$

In the case of a hadronic system the particle numbers $\{ N_1,...\:N_k \}$ 
are not conserved. Nevertheless the free energy $F(N_1,...\:N_k,V,T)$ with 
the numbers of particles $\{ N_1,...\:N_k \}$ being fixed is still of physical 
meaning. However, this is $
F(\langle N_1 \rangle ,...\:\langle N_k \rangle ,V,T)$, 
which corresponds to the thermodynamical (thermal and chemical) equilibrium.
The average particle numbers 
$\{ \langle N_1 \rangle ,...\: \langle N_k \rangle \}$ are found as a 
minimum of $F(N_1,...\:N_k,V,T)$. If the particles carry charges, say $+$ 
and $-$, and the total charge $Q$ is conserved, then the average values 
$\langle N^+ \rangle$ and $\langle N^- \rangle$ are also found as a minimum 
of $F$ but the additional constraint 
$\langle N^+ \rangle - \langle N^- \rangle = Q$ must be imposed. 

When we deal with the particles obeying quantum statistics, the approach 
described above, where the average particle numbers are found as a minimum 
of $F(N_1,...\:N_k,V,T)$ is rather inconvenient because there is no compact 
expression for $F(N_1,...\:N_k,V,T)$ even for the ideal gas. Then, we 
introduce the chemical potentials. However, this is a technical problem 
and not the matter of principles. 

After this comment we expand the free energy around the average particle 
numbers and take into account only the first three terms i.e.
\begin{eqnarray}
F(N_1,...\:N_k,V,T) &\cong& 
F(\langle N_1 \rangle,...\:\langle N_k \rangle,V,T) \\[4mm] \nonumber
&+& \sum_i {\partial F(N_1,...\:N_k,V,T) \over \partial N_i}
\Bigg\vert_{N_1=\langle N_1 \rangle,...\:N_k =\langle N_k \rangle}
(N_i -\langle N_i \rangle) \\[4mm] \nonumber
&+&{1 \over 2} \sum_{i,j} 
{\partial^2 F(N_1,...\:N_k,V,T) \over \partial N_i\partial N_j}
\Bigg\vert_{N_1=\langle N_1 \rangle,...\:N_k =\langle N_k \rangle}
(N_i -\langle N_i \rangle)(N_j -\langle N_j \rangle) \;.
\end{eqnarray}
Due to the relations
$$
\mu_i \buildrel \rm def \over = 
{\partial F(N_1,...\:N_k,V,T) \over \partial N_i}
\Bigg\vert_{N_1=\langle N_1 \rangle,...\:N_k =\langle N_k \rangle}
\;\;\;\;\;\;\;\;\;\;\;\; {\rm and}\;\;\;\;\;\;\;\;\;\;\;\; 
z_i \equiv e^{\beta \mu_i} \;,
$$
where $\mu_i$ is the chemical potential of $i-$the component, one finds
\begin{equation}\label{distribution}
P_{N_1,...\:N_k} \sim  
{\rm exp}\Bigg[-{1\over 2} \sum_{i,j} \Lambda_{ij}
(N_i -\langle N_i \rangle)(N_j -\langle N_j \rangle) \Bigg] \;,
\end{equation}
with the matrix $\Lambda$ given as
\begin{equation}\label{Lambda}
\Lambda_{ij} = \beta \;
{\partial^2 F(N_1,...\:N_k,V,T) \over \partial N_i\partial N_j}
\Bigg\vert_{N_1=\langle N_1 \rangle,...\:N_k =\langle N_k \rangle} \;.
\end{equation}
As well known, see e.g. \cite{Kor68}, the moment matrix $M_{ij}$ 
of the normal distribution (\ref{distribution}) equals
\begin{equation}\label{M}
M_{ij} \equiv 
\langle(N_i -\langle N_i \rangle)(N_j -\langle N_j \rangle) \rangle
= (\Lambda^{-1})_{ij} \;.
\end{equation}

We are now going to relate the matrix $\Lambda$ to the compressibility.
Since the free energy is an extensive quantity, which can be expressed as 
$$
F(N_1,...\:N_k,V,T) = V \, f(\rho_1,...\:\rho_1,T) \;,
$$
with $\rho_i \equiv N_i/V$, one shows that
\begin{eqnarray*}
p(N_1,...\:N_k,V,T) \buildrel \rm def \over = 
&-& {\partial F(N_1,...\:N_k,V,T) \over \partial V} \\[4mm]
= &-& {F(N_1,...\:N_k,V,T) \over V} + 
\sum_i \rho_i {\partial F(N_1,...\:N_k,V,T) \over \partial N_i}
\end{eqnarray*}
and then
\begin{equation}\label{ident}
- {\partial p(N_1,...\:N_k,V,T) \over \partial V}
= \sum_{i,j} \rho_i \rho_j 
{\partial^2 F(N_1,...\:N_k,V,T) \over \partial N_i \partial N_j} \;.
\end{equation}

Using eqs.~(\ref{M}) and (\ref{ident}) we get the final result
\begin{equation}\label{final1}
\sum_{i,j} \langle N_i \rangle \langle N_j \rangle (M^{-1})_{ij}  
=  - \; {V^2 \over T} \; \bigg({\partial p \over \partial V}
\bigg)_{T,\langle N \rangle_1,...\:\langle N \rangle_k,}  \;.
\end{equation}
The relation (\ref{final1}) essentially simplifies when the system 
components are independent from each other. Then, the matrices $\Lambda$ 
(\ref{Lambda}) and $M$ (\ref{M}) are diagonal ($M_{ij} = \Lambda_{ij} = 0$ 
for $i\not= j$) and 
\begin{equation}\label{final2}
\sum_i {\langle N_i \rangle^2 \over 
\langle N_i^2 \rangle - \langle N_i \rangle^2 }  
= - \; {V^2 \over T} \; \bigg({\partial p \over \partial V}
\bigg)_{T,\langle N \rangle_1,...\:\langle N \rangle_k,} \;.
\end{equation}
The question whether $\langle(N_i -\langle N_i \rangle)
(N_j -\langle N_j \rangle) \rangle = 0$ for $i\not= j$ can be answered
experimentally.

As already mentioned it is somewhat unusual in the statistical hadron 
physics to consider the thermodynamic quantities, like the compressibility 
from eqs.~(\ref{final1},\ref{final2}), at the fixed hadron average numbers. 
Computation of such quantities is straightforward if the hadrons are assumed 
to obey the Boltzmann statistics. In fact, this is a quite good approximation 
at the freeze-out stage when the hadron gas is expected to be rather dilute. 

We consider as an example the classical multi-component van der Waals gas. 
The equation of state is taken in the form
$$
p\:\Big( V - \sum_i \langle N_i \rangle v_i \Big) 
=  \sum_j \langle N_j \rangle T \;,
$$ 
where the parameter $v_i$ is related the volume of the $i-$sort hadron. 
The r.h.s. of eq.~(\ref{final1}) or (\ref{final2}) then equals
\begin{equation}\label{waals}
- \; {V^2 \over T} \; \bigg({\partial p \over \partial V}
\bigg)_{T,\langle N \rangle_1,...\:\langle N \rangle_k,}
= {\sum_j \langle N_j \rangle \over 
\Big(1 - \sum_i {\langle N_i \rangle v_i \over V} \Big)^2}
\cong \bigg(1 + 2 \sum_j {\langle N_i \rangle v_i \over V} \bigg)
\sum_j \langle N_j \rangle \;,
\end{equation}
where the last approximate equality assumes smallness of the
van der Waals correction. When $v_i \rightarrow 0$ we get the ideal
gas limit. The compressibility of the classical ideal gas is obtained 
from the relation (\ref{final1}) if the moment matrix $M$ (\ref{M}) 
equals $M_{ij} = \delta^{ij} \langle N_i \rangle$.

When the particles obey quantum statistics one usually introduces
the chemical potentials to compute the thermodynamical quantities.
Then, the pressure is a function of $\{\mu_1,.....\:\mu_k\}$ (not of 
$\{ \langle N_1 \rangle ,...\: \langle N_k \rangle \}$) and we express 
the derivative at fixed 
$\{ \langle N_1 \rangle ,...\: \langle N_k \rangle \}$, which is present
in eqs.~(\ref{final1},\ref{final2}), through the derivative at fixed
$\{\mu_1,.....\:\mu_k\}$. This is done by means of the respective jacobians,
see e.g. \cite{Lan80}. We give here the formula for a single component
system
$$
\bigg({\partial p \over \partial V}
\bigg)_{T,\langle N \rangle}
= \bigg({\partial p \over \partial V} \bigg)_{T,\mu} - 
{ \Big({\partial p \over \partial \mu} \Big)_{T,V} 
  \Big({\partial \langle N \rangle \over \partial V} \Big)_{T,\mu} 
\over 
\Big({\partial \langle N \rangle \over \partial \mu} \Big)_{T,V}} \;,
$$
and apply it to the ideal gas of pions which are bosons with three 
internal degrees of freedom. Then,
$$
\langle N(\mu,V,T) \rangle = 3V \int {d^3p \over (2\pi )^3} \;
\big( e^{\beta (E_p - \mu)} -1 \big)^{-1} \;,
$$
$$
p(\mu,V,T) = - 3T \int {d^3p \over (2\pi )^3} \; 
{\rm ln} \big( 1 - e^{- \beta (E_p - \mu)} \big) \;,
$$
where $E_p \equiv \sqrt{m^2 + {\bf p}^2}$ with $m$ being the pion
mass and ${\bf p}$ its momentum. For massless pions with
$\mu = 0$, which corresponds to the chemical equilibrium, we get
the r.h.s of eq.~({\ref{compres}) in the form
\begin{equation}\label{pions}
- {V^2 \over T} \;
\bigg({\partial p \over \partial V} \bigg)_{T,\langle N \rangle}
= {18 \zeta^2(3) \over \pi^4} \; VT^3 = 
{6 \zeta (3) \over \pi^2}\; \langle N \rangle 
\cong 0.71 \: \langle N \rangle \;,
\end{equation}
where $\zeta (z)$ is the Riemann function and $\zeta (3) \cong 1.202$. 
Comparing eqs.~(\ref{waals}) and (\ref{pions}) one sees that the particle 
number fluctuations are larger in the ideal gas of bosons than in the 
classical one.

Let us now discuss how to use eq.~(\ref{final1}) or (\ref{final2}) to
determine the hadron matter compressibility. The l.h.s. of these equations 
is fully determined by the particle multiplicity but to extract 
$\partial p /\partial V$ one also needs to deduce $T$ and $V$ in individual 
collisions. It is not a trivial task to get these three quantities.

One should first realize that the final state characteristics correspond 
to the so-called thermal freeze-out when the whole system disintegrates 
due to the switching-off the inter-particle interactions. However, the system 
chemical composition is fixed earlier, at the so-called chemical freeze-out 
when the interactions, which change the particle numbers, are no longer 
effective. The particle multiplicity, the system volume and temperature, 
which enter eqs.~(\ref{final1},\ref{final2}) refer to the chemical 
freeze-out. Therefore, these quantities, which are obtained from the
final state characteristics and consequently correspond to the thermal 
freeze-out, should be further recalculated to get their values at the 
chemical freeze-out. One can refer here to the procedure described in e.g. 
\cite{Bec97}, where the system volume and temperature at the chemical 
freeze-out are inferred from the multiplicities of the different sort 
particles.

The second but related difficulty lies in the fact that a sizeable fraction
of the final state particles originate from the decays of hadron resonances.
Since the particle multiplicity from eqs.~(\ref{final1},\ref{final2}) 
is that at the chemical freeze-out, the final state multiplicity should 
be recalculated to reconstruct the resonance contribution. This can be achieved 
within the thermodynamical model which takes into account hadron resonances, 
see e.g. \cite{Bra95,Bec97}.

The system temperature at the thermal freeze-out is usually found as an 
inverse slope of the transverse mass spectrum ($m_T = \sqrt{m^2 + p_T^2}$ 
with $p_T$ being the particle transverse momentum). This is however the 
`effective' temperature which incorporates the effect of transverse 
collective motion of the hadronic matter \cite{Lee90} and that of the 
resonance decays \cite{Sol91}. The actual temperature can be disentangled 
by the simultaneous analysis of different sort particles \cite{Lee90,Sol91}. 
The procedure is not very exact but the temperature is presumably measurable 
within 10\% accuracy.

The system volume in nuclear reactions is controlled due to the collision
centrality selection. In this way one can change the volume in a rather
broad range. One usually deduces the particle source size at the thermal 
freeze-out by means of the particle interferometry measurements, see e.g. 
\cite{Boa90}. The high multiplicity of relativistic nuclear collisions allows 
one for the event-by-event interferometry analysis. However, one should keep 
in mind that the hadron system decay is a dynamic process and the concept of 
the system volume is then not very precise. The hadron resonances also
complicate the analysis, see e.g. \cite{Wie97}. In spite of these 
difficulties, we still believe that the system volume can be measured with 
an accuracy which enables one to extract the hadron matter compressibility 
from eq.~(\ref{final1}).

How the data should be collected and processed? First of all one should
eliminate or at least reduce the trivial fluctuations due to the impact
parameter variation. This can be achieved by the trigger condition.
The collision centrality is well known to be strongly correlated to
the transverse or forward energy observed in the collision. Therefore,
selecting the transverse or/and forward energy from a narrow interval
we can collect a sample of collisions of a similar geometry. Then, 
one determines the system volume, temperature and hadron multiplicity
for every event. The events are split into subsamples in such a way 
that a subsample contains the events with coinciding or close temperatures 
and volumes. For every subsample one constructs the multiplicity distribution 
and computes the l.h.s. of eq.~(\ref{final1}) or (\ref{final2}). Then, one 
gets the compressibility for a given $V$ and $T$. Combining the results 
obtained for several subsamples we get the compressibility as a function 
of volume and temperature. The range of $V$ and $T$ will be presumably 
rather small. To get $\partial p /\partial V$ in a broader domain one
should vary the collision geometry by means of the trigger, change the
projectile-target system and the collision energy.

In summary, we have proposed a method to determine the hadronic matter
compressibility due to the event-by-event analysis of the particle
multiplicity fluctuations in heavy-ion collisions. The thermodynamical
relation, which connects the two quantities in the multi-component system, 
has been derived. As an illustration we have considered the compressibility 
of the classical van der Waals gas and the quantum gas of massless pions. 
Since the method requires not only the measurement of the particle 
multiplicity but of the system volume and its temperature as well, the 
problems related to such measurements have been briefly discussed. Finally, 
the procedure of data analysis has been suggested.


I am very grateful to Marek Ga\' zdzicki, Mark I. Gorenstein and Edward
V. Shuryak for critical reading of the manuscript and fruitful discussions.


\end{document}